\documentclass[aps,amsfonts,amsmath,prd,preprint,nofootinbib]{revtex4}
\pdfoutput=1
\newcommand{\beq}{\begin{equation}}
\newcommand{\eeq}{\end{equation}}
\usepackage{graphicx}
\usepackage{natbib}
\usepackage[utf8]{inputenc}
\begin{document}

\title{$\mu$-distortion around stupendously large primordial black holes}
\author{Heling Deng}
\email{heling.deng@asu.edu}
\affiliation{Physics Department, Arizona State University, Tempe, AZ 85287, USA}

\begin{abstract}
In a variety of mechanisms generating primordial black holes, each
black hole is expected to form along with a surrounding underdense region that
roughly compensates the black hole mass. This region will propagate
outwards and expand as a shell at the speed of sound in the homogeneous background. Dissipation of the shell due to Silk damping could lead to
detectable $\mu$-distortion in the CMB spectrum. While the current
bound on the average $\mu$-distortion is $\left| \bar{\mu}\right|\lesssim10^{-4}$,
the standard $\Lambda$CDM model predicts $\left| \bar{\mu}\right|\sim10^{-8}$,
which could possibly be detected in future missions. It is shown in
this work that the non-observation of $\bar{\mu}$ beyond $\Lambda$CDM
can place a new upper bound on the density of supermassive primordial
black holes within the mass range $10^{6}M_{\odot}\lesssim M\lesssim10^{15}M_{\odot}$.
Furthermore, black holes with initial mass $M\gtrsim10^{12}M_{\odot}$
could leave a pointlike distortion with $\mu\gtrsim10^{-8}$ at an
angular scale $\sim 1^{\circ}$ in CMB, and its non-observation would
impose an even more stringent bound on the population of these stupendously large primordial
black holes.

\end{abstract}

\maketitle

\section{Introduction}

Primordial black holes (PBHs) are hypothetical black holes formed
in the early universe before any large scale structures and galaxies.
Unlike astrophysical black holes formed by the collapse
of dying stars, PBHs are speculated
to be formed from large perturbations during the radiation
era (typically within the first few seconds after inflation ends)
and can have a mass ranging from the Planck mass ($\sim10^{-5}\ \text{g})$
to many orders of magnitude larger than the solar mass ($M_{\odot}\sim10^{33}\ \text{g}$).
The reader is referred to ref. \cite{Carr:2020gox} for an up-to-date review on PBHs
in regard of the mechanisms of formation and the current observational
constraints on their population in our universe.

PBHs have drawn much attention recently because they could
be responsible for the discoveries announced by the LIGO/Virgo Collaboration \cite{Bird:2016dcv,Sasaki:2016jop,Clesse:2016vqa}.
In the past few years, LIGO/Virgo detected about 50 signals \cite{LIGOScientific:2018mvr,Abbott:2020niy}, most of which are believed to be gravitational waves from inspiraling
and merging black holes of mass $\mathcal{O}(10-100)M_{\odot}$. The origin of these black holes is so far unknown. Some
of them have masses larger than what one would expect
in stellar models. For example, the surprising event GW190521 involves a black hole of mass 85$M_\odot$ \cite{Abbott:2020tfl}, which lies within the ``pair instability mass gap'' \cite{Woosley:2016hmi,Belczynski:2016jno,Spera:2017fyx,Giacobbo:2017qhh}. Hence the detection of the LIGO/Virgo black holes could be a hint of the existence of PBHs.

PBHs could also provide an explanation for supermassive black holes at the
center of most galaxies \cite{LyndenBell:1969yx,Kormendy:1995er}. These black holes have masses of $\mathcal{O}(10^{6}-10^{10})M_{\odot}$,
and observations of quasars indicated that some of them were already present at high redshifts. For example, a black hole of mass $M\sim10^{9}M_{\odot}$
was discovered recently from the most distant quasar ever observed at redshift $z\approx7.642$ \cite{wang2021luminous}. The existence of these black holes
greatly challenges the conventional mass accretion model of astrophysical black holes, as it is unable to fully explain the significant growth from stellar seeds \cite{Haiman:2004ve}.
PBHs, on the other hand, can be a natural candidate of these supermassive objects because they could have a large mass when they were born \cite{Rubin:2001yw,Bean:2002kx,Duechting:2004dk,Carr:2018rid}.

Yet another fascinating possibility, which has been under extensive investigation for
a long time, is that PBHs constitute (a major part of) the dark matter.
Observational bounds imposed by microlensing, dynamical and astrophysical
effects are stringent in most of the mass range (see, e.g., refs. \cite{Carr:2020gox,Carr:2020xqk} and references therein), but the window $\mathcal{O}(10^{17}-10^{23})\text{g}$
is open, hence can still allow small PBHs to account for all dark matter.

In the present work, we shall discuss a possible physical effect accompanied
by the production of PBHs: $\mu$-distortion in the spectrum of the cosmic microwave
background (CMB) around each PBH. Such an effect can arise in a variety
of mechanisms of PBH formation that require the total energy excess
of the perturbation to (approximately) vanish as the spacetime
goes asymptotically flat FRW. This requirement is not only satisfied for PBHs
from, e.g., topological defects or phase transitions \cite{Hawking:1987bn,Polnarev:1988dh,Garriga:1992nm,Caldwell:1995fu,Hawking:1982ga,Rubin:2000dq,Khlopov:2004sc,Garriga:2015fdk}, but could also
be fulfilled in the most popular scenario, where PBHs are formed by mass
overdensities that collapse upon Hubble reentry after inflation. If this condition is met, there would be a compensating underdense
region surrounding each PBH. After black hole formation, the underdense region will propagate outwards as a shell at the speed of sound. If the black hole is sufficiently massive, the
corresponding sound shell can have a wave energy so large that its dissipation due to Silk damping would lead to detectable
$\mu$-distortion in the CMB spectrum.\footnote{Note that this effect is accompanied with the formation of each single black hole, and is different from the average $\mu$-distortion produced by the dissipation of the background density fluctuations \cite{Carr:1993aq,Kohri:2014lza,Nakama:2016kfq,Nakama:2017xvq,Kawasaki:2019iis,Atal:2020yic}.}

It is the task of this paper to estimate the magnitude of the distortion and to investigate how it could be used to constrain
the PBH density. As we will see, while the resulting constraints could be applied
to the mass range $M\gtrsim10^{6}M_{\odot},$ they would be particularly
interesting for PBHs with an initial mass $M\gtrsim10^{11}M_{\odot}$,
which we call ``stupendously large PBHs'' following ref. \cite{Carr:2020erq}. Our main results
are shown in fig. \ref{fig:mu}.

The rest of the paper is organized as follows. In section \ref{sec:-distortion-around-PBH}
we discuss the evolution of the sound shell surrounding each PBH,
how its dissipation by Silk damping can lead to $\mu$-distortion
in CMB, and how the distortion could impose constraints on PBHs with
the help of future observations. Conclusions are summarized and discussed
in section \ref{discussion}. We set $c=\hbar=G=k_{B}=1$ throughout the paper.

\section{$\mu$-distortion around PBHs\label{sec:-distortion-around-PBH}}

\begin{figure}
\includegraphics[scale=0.55]{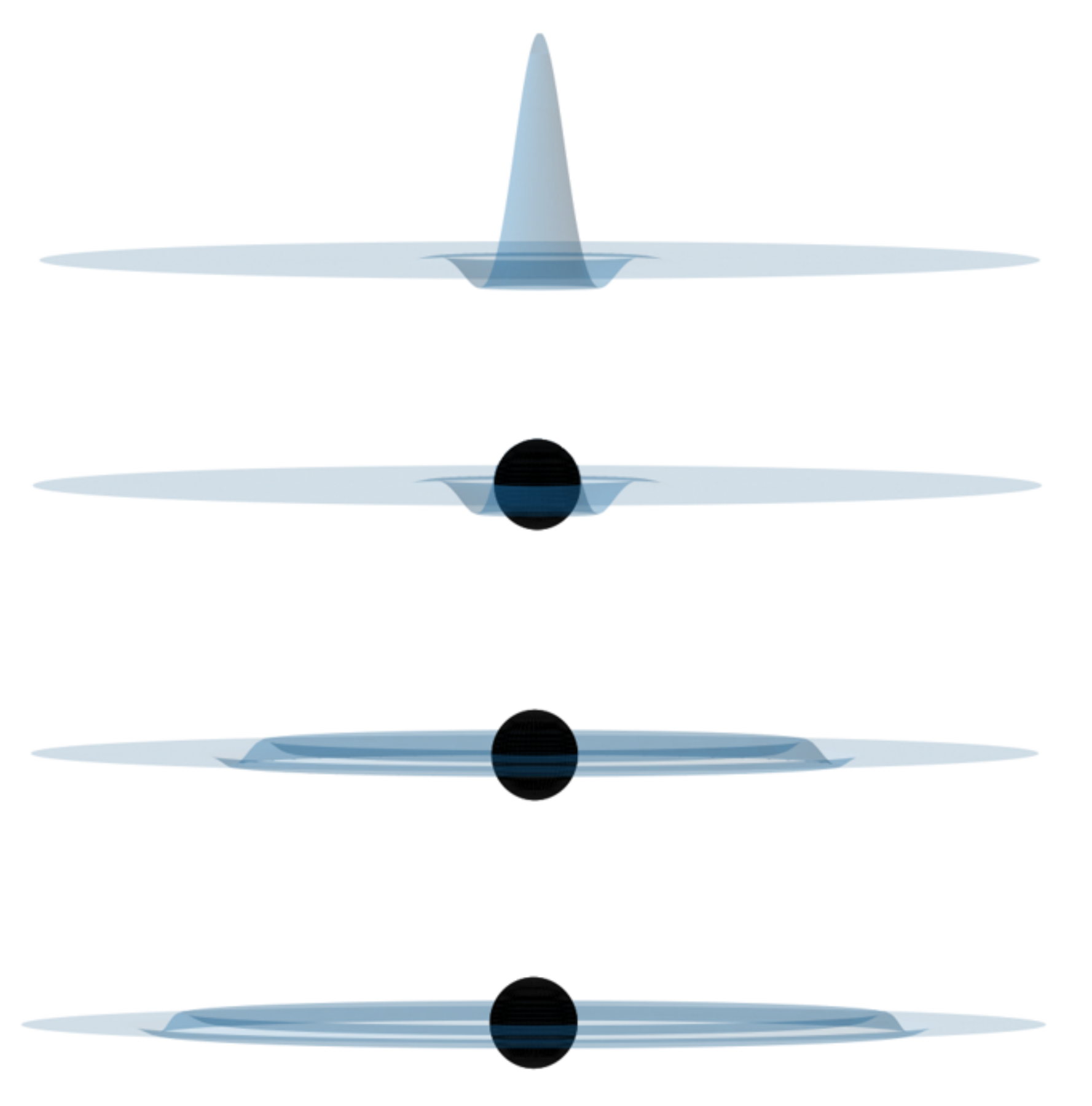}
\caption{\label{cartoon}Illustration of the formation of sound shell around a PBH. In this example the PBH is formed by the collapse a large overdense clump, which is surrounded by an underdense region. After the black hole is formed, the underdense region expands as a shell  (shown as a ring here). The shell consists of an underdense and an overdense layer, which is a typical feature of a spherical sound wave packet \cite{landau}. As the shell sweeps through, the fluid density between the black hole and the shell goes back to FRW.}
\end{figure}

In the most popular and perhaps the most natural scenario discussed in the literature, PBHs could
form by some rare overdense clumps in space that are from sufficiently large curvature perturbations at small scales. These perturbations can be attained by manipulating the inflationary potential \cite{Ivanov:1994pa,GarciaBellido:1996qt,Kawasaki:1997ju,Yokoyama:1998pt,Garcia-Bellido:2017mdw,Hertzberg:2017dkh}. After inflation ends, these clumps could overcome pressure and collapse into black holes as they reenter the Hubble horizon. In
the so-called ``compensated'' model \cite{Harada:2004pe,Carr:2014pga, Harada:2015yda}, the total energy excess of the perturbation is zero, and the spacetime away from the perturbation is flat FRW.\footnote{See also refs. \cite{Harada:2015yda, Musco:2018rwt} for discussion of the ``uncompensated'' model.} This implies there should be a compensating underdensity around the clump  (see, e.g., refs \cite{Germani:2018jgr,Kehagias:2019eil} for examples of the density profile from various power spectra of curvature perturbations). Therefore, each of the resulting black
holes is surrounded by an underdense region, whose energy deficit compensates
the black hole mass. This region will then propagate outwards as a
spherical sound wave packet, or, a sound shell. As the shell sweeps through the background, the fluid density between the black hole and the shell goes back to FRW. Illustrations of this process are shown in fig. \ref{cartoon}. Dissipation of the
shell due to photon diffusion during the so-called $\mu$-era can release energy into the background, generating $\mu$-distortion in the background photons, and may thus be seen in CMB. 

A similar picture can be applied to mechanisms of PBH formation involving
topological defects (such as cosmic strings and domain walls) or phase
transition bubbles \cite{Hawking:1987bn,Polnarev:1988dh,Garriga:1992nm,Caldwell:1995fu,Hawking:1982ga,Rubin:2000dq,Khlopov:2004sc,Garriga:2015fdk}. This is because, since
the global energy is conserved with the formation of these objects,
an energy excess that eventually turns into a black hole should be
compensated by a nearby energy deficit. We first noticed this effect
in our previous studies on PBHs formed by vacuum bubbles and spherical
domain walls that nucleate during inflation \cite{Deng:2016vzb,Deng:2017uwc,Deng:2020mds}, then in refs. \cite{Deng:2018cxb,Deng:2020pxo} we
further investigated the spectral distortions produced by the evolution
of the deficit in our specific models. It was later realized
that this effect can be generalized to a variety of PBH mechanisms,
which can be used to constrain the PBH density in general. 

\subsection{Dissipation of sound shell}

Before estimating the magnitude of the distortion, let us first describe
how the sound shell gets dissipated by photon diffusion in plasma.
More details can be found in the appendix. Roughly speaking, the black
hole is formed at a time $t_{M}$ when the Hubble horizon mass is
comparable to the black hole mass \cite{Carr:2020gox}, i.e.,
\begin{equation}
t_{M}\sim M.
\end{equation}
If the underdensity surrounding the black hole is of a scale
comparable to the black hole horizon ($2M$),
then since the total energy excess is assumed to vanish, the density 
contrast $\delta$ in the underdense region should satisfy $\mathcal{O}(10)\rho_{r}(t_{M})\delta M^{3}+M = 0$,
where $\rho_{r}(t)=3/32\pi t^{2}$ is the background FRW density.
This gives $\left|\delta\right|\sim 1$. As the underdensity propagates outwards in the form of a shell, $\left|\delta\right|$ becomes smaller and smaller, and the shell can thus be described as a spherical sound wave packet. According to hydrodynamics (and the appendix), the sound wave energy density is given by
$\rho_{s} =\rho_{r}\delta^{2}/4$,
hence the sound wave energy of the underdense shell $E_s(t)$ at $t_{M}$ is
$E_s(t_M) = \mathcal{O}(10)\rho_{s}(t_{M})M^{3}\sim M$.

If there is no photon diffusion, the thickness of the shell $D$ simply
gets stretched by cosmic expansion:
\begin{equation}
D(t)\sim\left(Mt\right)^{1/2},
\end{equation}
and the sound wave energy of the shell is $E_s(t)\sim M^{2}/D \propto t^{-1/2}$, which simply
gets redshifted over time. Due to photon diffusion, the photon-electron fluid can be effectively described by an imperfect fluid with a shear viscosity \cite{Weinberg:1971mx,weinberg2008cosmology,Pajer:2013oca}. By the appendix, the thickness of the shell $S$ gets further smeared, and can be
estimated as
\begin{equation}
S(t)\sim\left(D^{2}+\Lambda^{2}\right)^{1/2},\label{eq:S-1}
\end{equation}
where $\Lambda(t)$ is the physical scale of photon diffusion (the typical distance traveled by a photon till the time $t$), and is given by
\begin{equation}
\Lambda(t)\approx\left(\frac{t}{\sigma_{T}n_{e}}\right)^{1/2}\approx\left(\frac{t}{10^{10}\text{s}}\right)^{3/4}t^{1/2}\left(10^{12}M_{\odot}\right)^{1/2}.\label{eq:Lambda}
\end{equation}
Here, $\sigma_{T}$ is the Thomson cross-section and $n_{e}$ is the
electron density, and we have expressed $\Lambda$ in a form convenient
for our computations below.\footnote{To derive this expression, we have used the fact that the comoving scale of photon diffusion at $t=10^{10}\ \text{s}$ is about $ 0.14\ \text{Mpc}$ \cite{Khatri:2015tla}, and that the scale factor is $a(t)\approx (t/10^{19}\text{s})^{1/2}$.} By eq. (\ref{app:E_s}) in the appendix, the sound wave energy
of the shell becomes
\begin{equation}
E_{s}(t)\sim\left(\frac{D}{S}\right)^{3}\frac{M^{2}}{D}\sim\left[1+\left(\frac{t}{10^{10}\text{s}}\right)^{3/2}\left(\frac{M}{10^{12}M_{\odot}}\right)^{-1}\right]^{-3/2}M^{3/2}t^{-1/2},\label{eq:Es-1}
\end{equation}
which gets damped over time by a factor of $(D/S)^{3}$ compared to the case
without viscosity. Therefore, as the sound shell expands, part of
its wave energy is dumped into the space behind it, heating up the photons there. This energy release could lead to a unique
type of spectral distortion in CMB, which will be the topic of the next subsection.

\subsection{$\mu$-distortion}

CMB photons came from the last scattering surface with comoving
radius $\sim10\ \text{Gpc}$, at a time $\sim10^{12}-10^{13}\ \text{s}$
after inflation ends. During this time, photons decoupled from electrons and began to travel
across the transparent universe in all directions. Several generations
of detectors have shown with high precision that the CMB spectrum
is extremely close to a black-body of temperature $T\approx2.7\ \text{K}$. However,
tiny deviations from the black-body spectrum, commonly referred
to as spectral distortions, could arise from plenty of physical processes. Discovery of spectral distortions
would then provide valuable information of the pre-recombination universe.

A typical example of spectral distortions in CMB is known as the $\mu$-distortion, which is generated by the mix of photons
with different temperatures during the $\mu$-era, when $6\times10^{6}\ \text{s} \approx t_{th}\lesssim t\lesssim t_{\mu} \approx 10^{10}\ \text{s}$.
At time $t<t_{th}$, photons are completely
thermalized and have a perfect black-body spectrum. During the $\mu$-era,
photon number changing processes such as bremsstrahlung and double
Compton scattering become ineffective, while the Compton process still
allows photons to reach equilibrium with the background plasma. As a result, when there is energy released into the background, the
mix of photons with different temperatures would lead to a spectrum
with nonzero chemical potential $\mu$ \cite{sunyaev1970interaction,daly1991spectral,barrow1991primordial,Hu:1992dc}. 

Energy release in the early universe can occur in various possible mechanisms. A well known example is the Silk damping  \cite{Silk:1967kq}, where small-scale perturbations are smoothed out by photon diffusion before the time of recombination. In our scenario, the Silk damping of the sound shell constantly injects heat into the background as the shell
sweeps through space. During
the $\mu$-era, the comoving radius of
the shell (which is also approximately the comoving sound
horizon) ranges
from $r_{th}\sim0.1\ \text{Mpc}$ to $r_{\mu}\sim5\ \text{Mpc}$ . On the other hand, the comoving scale of photon diffusion
at the time of recombination is $\lambda_{rec}\sim50\ \text{Mpc}$. Because Silk damping continues to work after the $\mu$-era until recombination, $\lambda_{rec}\gg r_{\mu}$ means each black hole is at the center of a region of scale $\lambda_{rec}$
that has a distorted spectrum with a uniform $\mu$. Such a region will be referred to as
a ``Silk region'', and its projection on the last scattering surface will be called a ``Silk patch''. Let us now estimate the $\mu$-distortion around
a single black hole within a Silk region.

\begin{figure}
\includegraphics[scale=0.35]{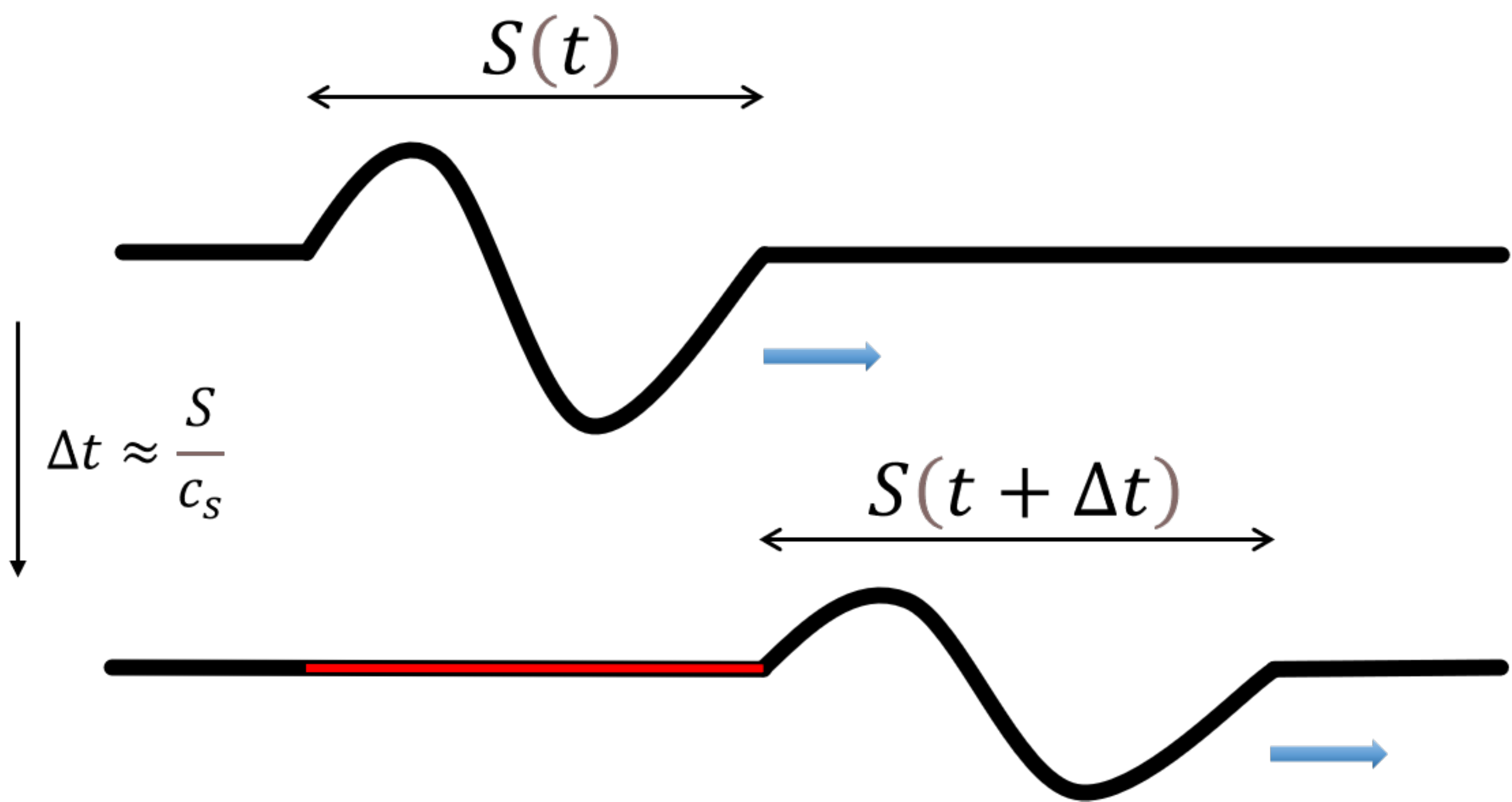}
\caption{\label{sketch}Sketch of the radiation fluid's density profile along the radius near the shell at two moments $t$ and $t+\Delta t$. The shell consists of an underdense and an overdense layer, which is a typical feature of a spherical sound wave packet \cite{landau}. The shell's physical radius is $\sim 2c_s t$, and its thickness increases from $S(t)$ to $S(t+\Delta t)$ due to cosmic expansion and the viscosity in fluid. The time it takes for the shell to completely pass through a sphere at the wave front is $\Delta t \sim S(t)/c_s$. During this time, the shell's wave energy gets dissipated, and the resulting heat is dumped behind the shell (shown as a red line).    }
\end{figure}

In the presence of energy release, the dimensionless chemical potential
$\mu$ ($\equiv-\mu_{th}/T$, where $\mu_{th}$ is the thermodynamic
chemical potential and $T$ is the background temperature) can be
obtained by \cite{sunyaev1970interaction,daly1991spectral,barrow1991primordial,Hu:1992dc}
\begin{equation}
\mu\approx1.4\frac{\Delta\rho_{s}}{\rho_{r}},
\end{equation}
where $\Delta\rho_{s}$ is the density of the released energy and
$\rho_{r}$ is the background radiation density. In calculating $\Delta\rho_{s}$ from $E_s$,
we ought to consider only the effect of dissipation, and not the dilution
from cosmic expansion. In the explicit expression of $E_s$ in eq. (\ref{eq:Es-1}), the effect of cosmic expansion is from $t^{-1/2}$. Therefore, as the sound shell sweeps through
a sphere at time $t$, the total released energy can be estimated as
\begin{equation}
\Delta E_{s}(t)\sim\frac{\text{d}\left(E_{s}t^{1/2}\right)}{\text{d}t}t^{-1/2}\Delta t,
\end{equation}
where $\Delta t\sim S/c_{s}$ is the time it take for the shell to
pass through the sphere ($c_s=1/\sqrt{3}$ is the speed of sound). Since the volume of the shell is $\sim 4\pi(2c_{s}t)^{2}S$, where $2c_{s}t$ is the shell's physical radius, we have
\begin{equation}
\Delta\rho_{s}(t)\sim\frac{\Delta E_{s}}{4\pi(2c_{s}t)^{2}S}.
\end{equation}
A sketch of this process is shown in fig. \ref{sketch}. Now the $\mu$-distortion produced at time $t$ can readily be expressed as
\begin{equation}
\mu(t)\approx4.85\frac{\text{d}\left(E_{s}t^{1/2}\right)}{\text{d}t}t^{-1/2}.\label{eq:mu}
\end{equation}

As the shell consecutively generates different values of $\mu$ 
behind it, photons of different spectra mix together. At the
time of recombination, the value of $\mu$ averaged over a Silk region
(of comoving volume $\sim\lambda_{rec}^{3}$) centered at a black hole of mass $M$ is 
\begin{align}
\mu_{M} & \sim\lambda_{rec}^{-3}\int_{\text{min}\{t_{th},t_{M}\}}^{t_{\mu}}\mu r^{2}\text{d}r=\lambda_{rec}^{-3}\int_{\text{min}\{t_{th},t_{M}\}}^{t_{\mu}}\mu r^{2}\dot{r}\text{d}t,\label{muM}
\end{align}
where $r\sim 2c_{s}t/a(t)$ is the comoving radius of the shell ($a(t)\sim (t/10^9\text{s})^{1/2}$ is the scale factor),
and $\text{min}\{t_{th},t_{M}\}$ denotes the smaller value between $t_{th}$
and $t_{M}$.\footnote{$\text{min}\{t_{th},t_{M}\}$ is set to be the lower bound in the integral because
for a black hole formed at $t_{M}>t_{th}\sim10^{12}M_{\odot}$, the
dissipation starts at $t_{M}$ instead of $t_{th}$. However, as we can see from the explicit expression
of $\mu_{M}$ below, for such a large black hole, $\mu_{M}$
is dominated by the production of $\mu$ close to $t_{\mu}$, hence
the lower bound is irrelevant.} By eqs. (\ref{eq:Es-1}), (\ref{eq:mu}) and (\ref{muM}),
$\mu_{M}$ can be found as
\begin{align}
\mu_{M} &\sim \lambda_{rec}^{-3}t_0^{3/2}\left.E_s(t)t^{1/2} \right|_{t_{\mu}}^{\text{min}\{t_{th},t_{M}\}}\\
& \sim10^{-25}\left(\frac{M}{M_{\odot}}\right)^{3/2}\left.\left[1+\left(\frac{t}{10^{10}\text{s}}\right)^{3/2}\left(\frac{M}{10^{12}M_{\odot}}\right)^{-1}\right]^{-3/2}\right|_{t_{\mu}}^{\text{min}\{t_{th},t_{M}\}}\label{eq:muM2}\\
 & \sim\begin{cases}
10^{-35}\left(\frac{M}{M_{\odot}}\right)^{3}, & M<10^{7}M_{\odot},\\
10^{-25}\left(\frac{M}{M_{\odot}}\right)^{3/2}, & 10^{7}M_{\odot}<M<10^{12}M_{\odot},\\
10^{-13}\left(\frac{M}{M_{\odot}}\right)^{1/2}, & 10^{12}M_{\odot}<M<10^{15}M_{\odot},
\end{cases}\label{eq:muM}
\end{align}
where in the first line, $t_0$ is the present time. Since black holes formed at $t_{\mu}\approx 10^{10}\ \text{s}$ have mass $M\sim10^{15}M_{\odot}$,
this expression cannot be applied to larger black holes. Eq. (\ref{eq:muM2}) or (\ref{eq:muM})
above gives the $\mu$-distortion in CMB produced by a single black
hole within a Silk region. If PBHs within the relevant mass range
are rare on the CMB sky, the distortions are pointlike. If there are
more than one of these PBHs within the Silk region, the resulting
$\mu$ should be the sum of contributions from all of them, and
we would have an average $\mu$ in the CMB spectrum, $\bar{\mu}$. As it turns out,
sufficiently massive PBHs with a sufficiently large density can
produce a detectable distortion. Inversely, the non-observation
of $\mu$-distortion is CMB can constrain the PBH density. This will be the topic of the next subsection.

\subsection{Potential observational constraints on PBHs}

Typically, CMB spectral distortions are expected to be isotropic. The current observational upper bound on the all-sky $\mu$-distortion, $\bar{\mu}$,
is an old result from the seminal COBE/FIRAS measurements, which gave $\left| \bar{\mu}\right|\lesssim10^{-4}$ \cite{Fixsen:1996nj}. Hypothetical future experiments under discussion are expected to significantly improve this limit, reaching $\left| \bar{\mu}\right|\sim 10^{-8}$ \cite{Kogut:2019vqh,Chluba:2019nxa}. This can be a test of the standard $\Lambda$CDM model, because by assuming a nearly scale-invariant spectrum of the primordial perturbations, the damping of small-scale acoustic modes results in $\bar{\mu}\sim10^{-8}$ \cite{Cabass:2016giw,Chluba:2016bvg}. The non-detection of this predicted distortion would thus inevitably require new physics beyond the vanilla slow-roll inflationary model. 

As mentioned at the end of last subsection, there are two kinds of $\mu$-distortion in our scenario: an all-sky $\bar{\mu}$, and a local $\mu_M$. A Silk patch is only a tiny part of the CMB sky. It
is possible that there are some rare spots containing only one large PBH accompanied by a detectable local distortion.  In order to receive signals of this kind, we need an image of CMB with a sufficiently high resolution. While the angular size of a Silk patch is $\delta \theta \sim 0.2^\circ$, hypothetical missions being proposed and discussed can only reach a resolution of $\delta \theta\lesssim 1^\circ$ \cite{Chluba:2019nxa}. A patch of this size contains $\mathcal{O}(10)$ Silk patches. Therefore, if there is only one PBH in this patch, the magnitude of the pointlike distortion should be reduced to $\sim 0.1\mu_M$. 

\begin{figure}
\includegraphics[scale=0.35]{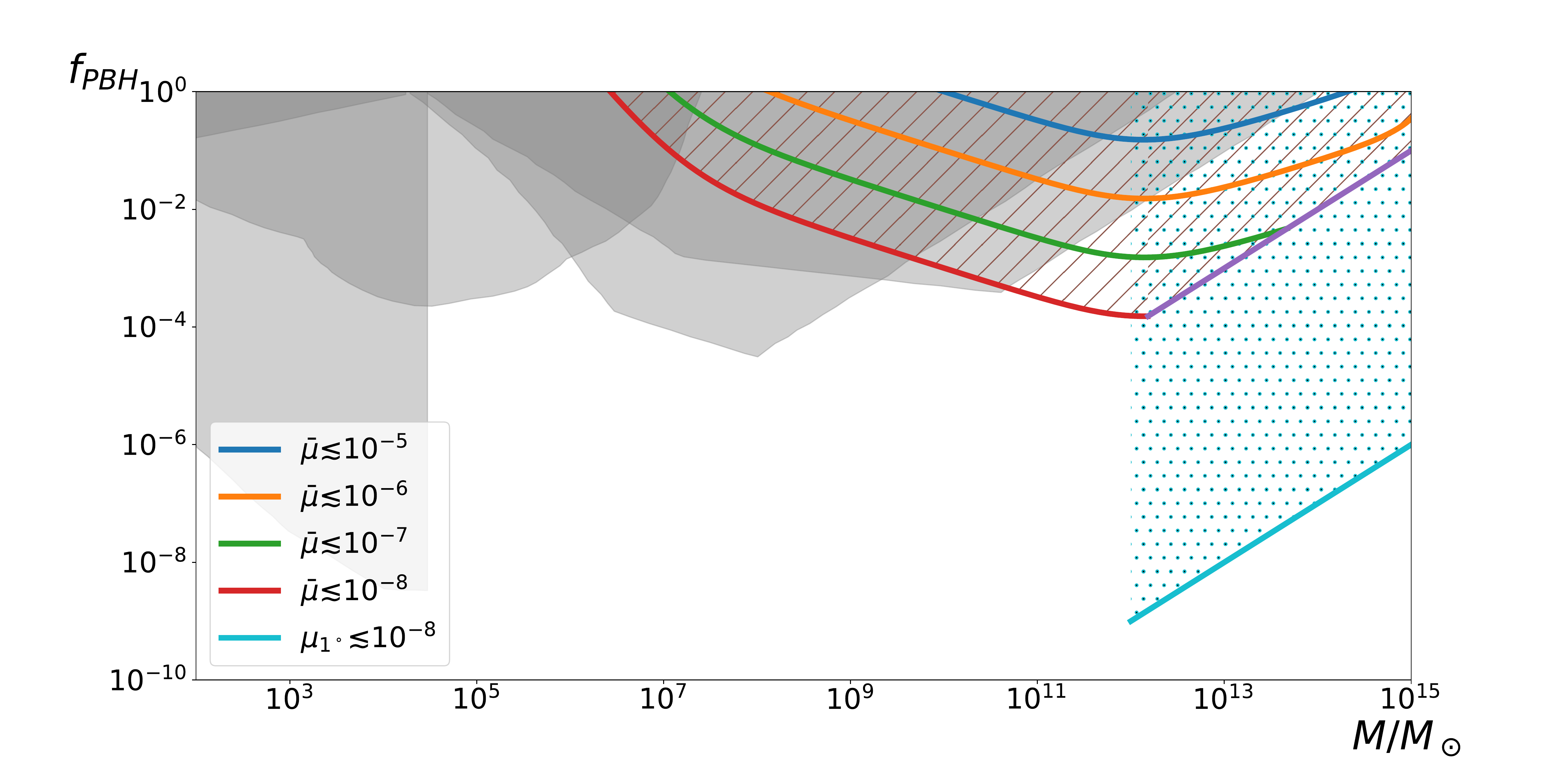}

\caption{\label{fig:mu}Constraints on the fraction of dark matter in monochromatic
PBHs within the mass range $10^2-10^{15}M_{\odot}$ . The gray regions (adapted from fig. 10 in ref. \cite{Carr:2020gox}) have been ruled out by current observations \cite{Oguri:2017ock,Serpico:2020ehh,Inoue:2017csr,Carr:2018rid}. Colored curves
are possible upper bounds for $f_{PBH}$ if future observations find certain upper
bounds for the average $\mu$-distortion in CMB: $\bar{\mu}\lesssim10^{-5}$ (blue),
$\bar{\mu}\lesssim10^{-6}$ (orange), $\bar{\mu}\lesssim10^{-7}$
(green) and $\bar{\mu}\lesssim10^{-8}$ (red). The purple line is imposed by condition (\ref{condition}) that there should at least be one black holes within the smallest patch future missions can measure. The
cyan line is a possible bound if future observations find $\mu\lesssim10^{-8}$
at an angular scale $\sim 1^{\circ}$.}
\end{figure}

Let us now estimate how $\bar{\mu}$ and $\mu_M$ in our scenario can be used to constrain the PBH density. Let $f(M)$ be the fraction of dark matter in PBHs
(which is sometimes called the PBH mass function) defined by
\begin{equation}
f(M)\equiv \frac{M}{\rho_{DM}}\frac{\text{d}n}{\text{d}\ln M},
\end{equation}
where $\text{d}n$ is the number density of black holes within the mass range ($M,M+\text{d}M$), and $\rho_{DM}$ is the energy density of dark matter. Then the comoving number density of PBHs of mass
$\sim M$ can be estimated as
\begin{equation}
n_{M}\sim\rho_{DM}(t_{0})\frac{f}{M}\approx2.8\times10^{10}\frac{M_{\odot}}{M}f\ \text{Mpc}^{-3}\sim 10^{15}\frac{M_{\odot}}{M}f\lambda_{rec}^{-3},\label{eq:nM}
\end{equation}
where $t_0$ is the present time. Therefore, the total number of PBHs with mass $\sim M$ within a Silk region
is $\sim n_{M}\lambda_{rec}^{3}\sim\left(10^{15}M_{\odot}/M\right)f$. In order to have an all-sky $\mu$, we should have at least one PBH within the patch of angular scale $\delta \theta\sim 1^\circ$, which contains $\mathcal{O}(10)$ Silk patches. We thus need
\begin{equation}
\mathcal{O}(10) n_{M}\lambda_{rec}^{3}\sim 10^{16}\frac{M_{\odot}}{M}f \gtrsim 1. \label{condition}
\end{equation}
When this condition is satisfied, the all-sky $\mu$ is
\begin{equation}
\bar{\mu}\sim \mu_M n_{M}\lambda_{rec}^{3},\label{mubar}
\end{equation}
where $\mu_M$ is given by eq. (\ref{eq:muM2}) or (\ref{eq:muM}). The current bound $\bar{\mu}\lesssim 10^{-4}$ does not place any constraints on $f$, as it gives $f<1$ for
any $M$.  However, the non-observation
of $\bar{\mu}>10^{-8}$ in future missions can constrain the density of supermassive PBHs. By assuming $\bar{\mu}\lesssim10^{-8}$ in eq. (\ref{mubar}),
we find
\begin{equation}
f\lesssim\begin{cases}
10^{12}\left(\frac{M}{M_{\odot}}\right)^{-2}, & M<10^{7}M_{\odot},\\
10^{2}\left(\frac{M}{M_{\odot}}\right)^{-1/2}, & 10^{7}M_{\odot}<M<10^{12}M_{\odot},\\
10^{-10}\left(\frac{M}{M_{\odot}}\right)^{1/2}, & 10^{12}M_\odot < M < 10^{15}M_{\odot},
\end{cases}\label{f}
\end{equation}
which constrains the PBH density for $M\gtrsim 10^{6}M_{\odot}$. On the other hand, the necessary condition (\ref{condition}) gives
\begin{equation}
f \gtrsim 10^{-16}\frac{M}{M_{\odot}},\label{f2}
\end{equation}
which is required in order for the bound (\ref{f}) to be valid. We show this possible
bound as the red and purple lines in fig. \ref{fig:mu}, along with other
bounds assuming $\bar{\mu}\lesssim10^{-5},10^{-6},\ \text{and}\ 10^{-7}$. Note
that instead of plotting them as broken power laws as in (\ref{f}),
we show in fig. \ref{fig:mu} the bounds with smooth functions
derived from eq. (\ref{eq:muM2}).

As for the spiky distortions in possibly rare spots on the CMB sky, by eq. (\ref{eq:muM}) we have $\mu_M>10^{-7}$ for
$M>10^{12}M_{\odot}$. Considering that signals of this kind will be measured in a patch of angular scale $\sim 1^\circ$, which contains $\mathcal{O}(10)$ Silk patches, a single black hole with $M>10^{12}M_{\odot}$ should give $\mu_{1^\circ}>10^{-8}$ in a single pixel. If there is no indication of such signals, the PBH mass function can be constrained
by $n_{M}\lesssim \left[4\pi\left(10\ \text{Gpc}\right)^{2}\left(50\ \text{Mpc}\right)\right]^{-1} \sim 10^{-11}\ \text{Mpc}^{-3}$,
which means there is not a single stupendously large black hole living
on the CMB sky (here $10\ \text{Gpc}$ is the comoving radius of the
last scattering surface). By eq. (\ref{eq:nM}), we have
\begin{equation}
f(M>10^{12}M_{\odot})\sim10^{-10}\frac{M}{M_{\odot}}n_{M}\ \text{Mpc}^3\lesssim 10^{-21}\frac{M}{M_{\odot}},
\end{equation}
which is even more stringent than the bound given by $\bar{\mu}$
at $M>10^{12}M_{\odot}$. This possible bound is shown as the cyan
line in fig. \ref{fig:mu}.

\section{Conclusions and discussion \label{discussion}}

In this paper we have studied the possible $\mu$-distortion in the CMB spectrum
around supermassive PBHs and how future observations might impose constraints
on the PBH density within the mass range $\mathcal{O}(10^{6}-10^{15})M_{\odot}$.
Our discussion is based on the assumption that each PBH is formed
with a surrounding underdense region that roughly compensates the black hole
mass. This should apply to a number of mechanisms of PBH
formation, including PBHs from inflationary density fluctuations,
topological defects, phase transitions, etc., where the total mass excess of the perturbation (approximately) vanishes. As the underdense
region expands outwards as a wave packet at the speed of sound, which
we refer to as a sound shell, dissipation of the wave due to photon
diffusion can inject energy into the background and thus produce $\mu$-distortion
in the CMB spectrum. 

If there are more than one PBH with $M\gtrsim10^{6}M_\odot$ within a Silk region, which is of an angular scale of $0.2^{\circ}$, we can have an average distortion
$\bar{\mu}$ in CMB. The possible non-observation of $\bar{\mu}$ in future missions
beyond the $\Lambda$CDM model (which predicts $\bar{\mu}\sim10^{-8}$)
would then place constraints on these black holes. A bound of particular
interest would be $f<10^{-3}$ for $10^{10}M_{\odot}\lesssim M\lesssim10^{12}M_{\odot}$. 

Even if these supermassive PBHs are rare in our universe, it is possible that we can see pointlike distortions with magnitude $\mu_M\gtrsim10^{-7}$ on some Silk patches in CMB, as long as the black holes have initial mass $M\gtrsim10^{12}M_{\odot}$. Considering that the resolution of future missions could reach $\delta \theta \sim 1^\circ$, such a signal in a pixel would be $\mu_{1^\circ}\gtrsim10^{-8}$. The non-observation would imply
that these stupendously large PBHs can only constitute a tiny part
of the dark matter, with a fraction $f<10^{-9}$ for $M\sim10^{12}M_{\odot}$. 

Our main results are shown in fig. \ref{fig:mu}. If future observations do not see $\mu$-distortion with $\mu\gtrsim 10^{-8}$, the shaded regions in the figure, including that with (brown) tilted lines and that with (cyan) dots, should all be excluded. An exciting possibility is that we do see local distortions in CMB, which would be a hint of the existence of stupendously large PBHs, and we will be able to estimate their population by eq. (\ref{eq:muM}).

Two assumptions that lead to our results are (a) the black hole
is formed at a time $t_{M}\sim M$;
(b) the underdense region is of a scale $D\sim M$ at $t_M$. While these are plausible assumptions, the exact values
of $t_{M}$ and $D$ are model dependent, and can even vary with different PBH masses in a given model. Typically, we would expect $t_{M}=\mathcal{O}(1-10)M$ and $D=\mathcal{O}(1-10)M$. In this case, the bounds shown in
fig. \ref{fig:mu} should be adjusted accordingly.

An important aspect we did not consider is the mass accretion of PBHs. It is possible that supermassive PBHs did not acquire much accretion,
which naturally explains the proportionality between the masses of
supermassive black holes and those of galaxies \cite{Carr:2018rid}. But this seems unlikely. Although the increase in mass is insignificant during the radiation
era, a large amount of gas (including dark matter), be it galactic
or intergalactic, is expected to fall into the black hole after the
dust-radiation equality. Many works have been done constraining the PBH density with mass accretion of the Bondi-type \cite{carr1981pregalactic,Ricotti:2007au,Ricotti:2007jk,Ali-Haimoud:2016mbv}. The accretion rate, however, is highly model
dependent. It is also known (and particularly pointed out in ref. \cite{Carr:2020erq}) that the Bondi process
is only valid for black holes with $M\lesssim10^{4}M_{\odot}$, and
the physics of mass accretion for stupendously large PBHs is still unclear. When taking into account mass accretion, a naive expectation is that the bounds shown
in fig. \ref{fig:mu} should be deformed and shifted toward the right.

Lastly, although the focus of the present work is on the $\mu$-distortion generated by the Silk damping of the sound shell, it should be mentioned that the shell itself can be a source of temperature perturbations in CMB. At the time of recombination ($t\sim 10^{13}\ \text{s}$), the shell's radius ($\sim$ sound horizon) is of an angular size $\sim 1^\circ$ on the CMB sky, and its thickness is the Silk damping scale $\sim 0.2^\circ$ ($S\sim \Lambda$), which is comparable to the thickness of the last scattering surface. Therefore, the underdensity and overdensity on the shell (fig. \ref{sketch}) can induce a ring-like feature in CMB (the ring's radius depends on the black hole's position relative to the last scattering surface). Since temperature in the radiation era is given by $T\propto \rho^{1/4}$, where $\rho$ is the radiation density, then by eq. (\ref{delta_am3}) in the appendix, the temperature perturbation can be estimated as
\begin{equation}
\left| \frac{\delta T}{T}\right| \approx \frac{1}{4}\delta_{am}\sim 10^{-26}\left(\frac{M}{M_{\odot}}\right)^{3/2}. 
\end{equation}
For a black hole as massive as $10^{14}M_\odot$, this gives $\sim 10^{-5}$, comparable to the observed rms CMB perturbations. Hence PBHs with larger initial masses should be rare on the last scattering surface, which means these PBHs are constrained by the cyan bound in fig. \ref{fig:mu}.

\section*{Appendix: Sound shell in a viscous radiation fluid}

In this appendix we study the evolution of a spherical sound wave packet expanding
in a radiation-dominated universe, and how its energy gets dissipated by viscosity. 

\subsection*{(a) Minkowski, without viscosity}

We first consider the sound wave equations for a perfect radiation fluid
in Minkowski spacetime. Let $\rho(x^{i},\tau)$ be the energy
density of the fluid, $\delta=\rho/\rho_{r}-1$ be the small density
contrast (where $\rho_{r}$ is the unperturbed background density), and $u^{i}$
be the fluid velocity. The energy-momentum tensor of the fluid is
\begin{align}
T^{00} & =\rho_{r}(1+\delta)+\frac{4}{3}\rho_{r}u^{2},\\
T^{i0} & =\frac{4}{3}\rho_{r}(1+\delta)u^{i},\\
T^{ij} & =\frac{1}{3}\delta^{ij}\rho_{r}(1+\delta)+\frac{4}{3}\rho_{r}(1+\delta)u^{i}u^{j}.
\end{align}
where $u^{2}=|u^{i}u_{i}|$, and we have taken into account that the radiation
pressure is $p=\rho/3$. The sound wave
equations can then be found by the lineraized energy-momentum conservation,
$\partial_{\mu}T^{\mu\nu}=0$:
\begin{align}
3\dot{\delta}+4\partial_{i}u^{i} & =0,\label{eq:sound1}\\
4\dot{u}^{i}+\partial_{i}\delta & =0,\label{eq:sound2}
\end{align}
where $\dot{}\equiv\partial/\partial\tau$. Now let $u^{i}\equiv\partial_{i}\phi$,
where $\phi$ is known as the velocity potential. Then by the second
equation, we have $\delta=-4\dot{\phi}$, and the two equations can
be combined into
\begin{equation}
\ddot{\phi}-\frac{1}{3}\nabla^{2}\phi=0,\label{eq:wave equation}
\end{equation}
which describes a wave propagating at the speed of sound $c_{s}=1/\sqrt{3}$.
For a spherical wave, this equation has a general solution, $\phi=f(r-c_{s}\tau)/r$. Furthermore, since in this work we are interested in a spherical sound shell, we will approximate the wave by writing its velocity potential in a Gaussian form:
\begin{equation}
\phi=K\frac{d^{2}}{r}e^{-\frac{(r-r_{s})^{2}}{2d^{2}}},\label{eq:phi}
\end{equation}
where $d$ is the shell's thickness, $r_{s}\equiv c_{s}\tau$ is the
shell's radius, and $K$ is a dimensionless factor. We will also assume the shell's thickness is negligible compared
to its radius, i.e., $r_{s}\gg d$. Then the density contrast and radial
fluid velocity are found to be
\begin{equation}
\delta=-4c_{s}K\left(1-\frac{r_{s}}{r}\right)e^{-\frac{(r-r_{s})^{2}}{2d^{2}}},
\end{equation}
\begin{equation}
u=-K\left(1+\frac{d^{2}}{r^{2}}-\frac{r_{s}}{r}\right)e^{-\frac{(r-r_{s})^{2}}{2d^{2}}}\approx-K\left(1-\frac{r_{s}}{r}\right)e^{-\frac{(r-r_{s})^{2}}{2d^{2}}}=\delta/4c_{s}.
\end{equation}
We can see that $\delta>0$ for $r<r_{s}$ and $\delta<0$ for $r>r_{s}$,
which means the shell consists of an underdensity and an overdensity. This is a typical feature of a spherical sound wave packet \cite{landau}. The amplitude of $\delta$ can be estimated as the value of $|\delta|$ at $r=r_s+d$, which gives
\begin{equation}
\delta_{am} \sim K\frac{d}{r_s}. \label{delta_am}
\end{equation}

The energy density of the fluid is
\begin{equation}
T^{00}=\rho_{r}+\rho_{r}\delta+\frac{4}{3}\rho_{r}u^{2},
\end{equation}
where the three terms are respectively the densities of the background, the deficit and the sound wave energy. Therefore, the energy deficit and the sound
wave energy of the shell are given by
\begin{equation}
E_{-}=4\pi\rho_{r}\int\delta r^{2}dr\approx-73K\rho_{r}d^{3},\label{eq:E-}
\end{equation}
\begin{equation}
E_{s}=4\pi\rho_{r}\int\frac{4}{3}u^{2}r^{2}dr\approx21K\rho_{r}d^{3},\label{eq:Es}
\end{equation}
both of which are constant over time. 

\subsection*{(b) Minkowski, with viscosity}

In the presence of photon diffusion, the energy-momentum tensor of
the fluid acquires an extra term $\Delta T_{\mu\nu}$ that acts effectively
as viscosity \cite{Weinberg:1971mx,weinberg2008cosmology,Pajer:2013oca}. For an irrotational flow its divergence is given by
\begin{align}
\partial_{\nu}\Delta T^{0\nu} & =0,\\
\partial_{\nu}\Delta T^{i\nu} & =-\frac{4}{3}\eta\nabla^{2}u^{i},
\end{align}
where the viscous coefficient $\eta$ is
\begin{equation}
\eta=\frac{16}{45}\rho_{r}\tau_{\gamma}.
\end{equation}
Here, $\tau_{\gamma}=(\sigma_{T}n_{e})^{-1}$ is the photon's mean
free path, where $\sigma_{T}$ is the Thomson cross-section and $n_{e}$
is the electron density. With this extra term, eq. (\ref{eq:sound2})
becomes
\begin{equation}
4\dot{u}^{i}+\partial_{i}\delta-\frac{4\eta}{\rho_{r}}\partial^{2}u^{i}=0,
\end{equation}
while eq. (\ref{eq:sound1}) remains unchanged. Then the wave equation
(\ref{eq:wave equation}) becomes
\begin{equation}
\ddot{\phi}-\frac{1}{3}\nabla^{2}\phi-\frac{\eta}{\rho_{r}}\nabla^{2}\dot{\phi}=0.
\end{equation}
In the limit of small viscosity, this can be solved in the Fourier
representation, with the result
\begin{equation}
\phi=K\left(\frac{d}{s}\right)\frac{d^{2}}{r}e^{-\frac{(r-r_{s})^{2}}{2s^{2}}},
\end{equation}
where
\begin{equation}
s=\sqrt{d^{2}+\lambda^{2}}
\end{equation}
with $\lambda\equiv\left(\eta t/\rho_{r}\right)^{1/2}$. By the definition
of $\eta$, $\lambda\sim\left(\tau_{\gamma}t\right)^{1/2}$, which
is the photon diffusion scale. Comparing this solution with eq. (\ref{eq:phi}),
we can see that the shell's thickness increases with time due to the
viscosity, while the amplitude of the velocity potential gets damped
by $d/s$. 

The fluid velocity $u=\partial_{r}\phi$ and density contrast $\delta=-4\dot{\phi}$ are then given by
\begin{equation}
u\approx \frac{\delta}{4c_s} \approx-K\left(\frac{d}{s}\right)^{3}\left(1-\frac{r_{s}}{r}\right)e^{-\frac{(r-r_{s})^{2}}{2s^{2}}}.\label{eq:u}
\end{equation}
As before, the amplitude of $\delta$ can be estimated as the value of $|\delta|$ at $r=r_s+s$, which is
\begin{equation}
\delta_{am} \sim K\left(\frac{d}{s}\right)^{2}\frac{d}{r_s}. \label{delta_am2}
\end{equation} 

The energy deficit and sound wave energy of the
shell now become
\begin{equation}
E_{-}\approx -73K\rho_{r}d^{3}.
\end{equation}
\begin{equation}
E_{s}\approx21K\left(\frac{d}{s}\right)^{3}\rho_{r}d^{3}.
\end{equation}
We can see $E_s$ gets damped by $(d/s)^3$ due to the viscosity, whereas $E_{-}$ remains constant over time.

\subsection*{(c) Flat FRW, with viscosity}

Now if we take into account cosmic expansion, the spatial coordinate
$r$ becomes the comoving radius, and time $\tau$ is understood as
the conformal time, i.e., $a\text{d}\tau=\text{d}t$, where $t$ is
the physical time and $a$ is the scale factor. In a flat, radiation-dominated
universe, we have $a(\tau)=a_{m}\tau$ and $t=a_{m}\tau^{2}/2$, where
$a_{m}$ is an integration constant \cite{mukhanov2005physical}. Then eq. (\ref{eq:u}) becomes
\begin{equation}
u\approx \frac{\delta}{4c_s} \approx -K\left(\frac{D}{S}\right)^{3}\left(1-\frac{R_{s}}{R}\right)e^{-\frac{(R-R_{s})^{2}}{2S^{2}}}
\end{equation}
where $R$ is the physical spatial coordinate, $R_{s}=a(\tau)c_{s}\tau=2c_{s}t$
and $S=as$ are respectively the physical radius and thickness of the sound shell, and $D=ad$. The amplitude of $\delta$ can be estimated as
\begin{equation}
\delta_{am} \sim K\left(\frac{D}{S}\right)^{2}\frac{D}{R_s}. \label{delta_am3}
\end{equation}

The sound wave energy of the
shell now becomes
\begin{equation}
E_{s}(t)\approx21K\left(\frac{D}{S}\right)^{3}\rho_{r}D^{3}.\label{eq:EsD}
\end{equation}
Without the damping factor $(D/S)^3$, we have $D=S,$ then since $\rho_{r}\propto a^{-4}$
and $D^{3}\propto a^{3}$, $E_{s}$ gets redshifted over time. 

In our scenario, the sound shell starts as an underdense region compensating the black hole mass $M$ when  the black hole is formed. By the results in eqs.
(\ref{eq:E-}) and (\ref{eq:Es}), this means $E_{s}(t_M)\sim M$, where $t_M$ is the black hole formation time. For simplicity, we
assume $t_{M}\sim D(t_{M})\sim M$, which then gives
\begin{equation}
E_{s}(t)\sim\left(\frac{D}{S}\right)^{3}\frac{M^{2}}{D}. \label{app:E_s}
\end{equation}
Here, $D\sim\sqrt{Mt}$ and $S=\sqrt{D^{2}+\Lambda^{2}}$, 
with $\Lambda=a\lambda$ being the physical scale of photon diffusion. 

It can also be shown that the amplitude of the density contrast $\delta$ can be rewritten as
\begin{equation}
\delta_{am} \sim \left(\frac{D}{S}\right)^2\frac{M}{D},
\end{equation}
which can be used to estimate the temperature perturbations on the shell.

\section*{Acknowledgments}
I am grateful to Alex Vilenkin for stimulating discussions, which led to this work. I would also like to thank Tanmay Vachaspati for insightful comments on the manuscript. This work is supported by the U.S. Department of Energy, Office of High Energy Physics, under Award
No. de-sc0019470 at Arizona State University.

\bibliography{mulib}
\end{document}